\documentclass[pra,aps,longbibliography,showpacs,groupedaddress,superscriptaddress,floatfix,twocolumn,toc=flat]{revtex4-1}

\usepackage{amsfonts,amsmath,amssymb,stmaryrd}

\usepackage[utf8]{inputenc}
\usepackage[table]{xcolor}
\usepackage{bbold, bm}
\usepackage{graphicx}
\usepackage{array,multirow}
\usepackage{algorithm}
\usepackage{algpseudocode}
\algrenewcommand\algorithmicrequire{\textbf{Input:}}
\algrenewcommand\algorithmicensure{\textbf{Output:}}
\usepackage{boldline}
\usepackage{tabularx}
\usepackage{tikz}
\usepackage{placeins}
\usepackage{mathrsfs}
\pdfmapline{-dummy LMRoman10-Regular} 
\pdfmapline{-dummy LMRoman10-Bold} 
\pdfmapline{-dummy LMRoman10-Italic} 
\pdfmapline{-dummy LMRoman10-BoldItalic} 
\pdfmapline{-dummy LMRoman12-Regular} 
\pdfmapline{-dummy LMRoman12-Bold} 
\pdfmapline{-dummy LMRoman12-Italic} 
\pdfmapline{-dummy LMRoman12-BoldItalic} 

\usepackage{ctable}
\usepackage{epsfig}
\usepackage[english, status=final]{fixme}
\fxsetup{marginface=\tiny}
\fxusetheme{color}    

\usepackage[colorlinks=true,linkcolor=blue,urlcolor=blue,citecolor=blue]{hyperref}

\newcommand{\G}{\hat{G}}
\newcommand{\Gj}{\hat{G}_{\bm{j}}}

\newcommand{\ket}[1]{|#1\rangle}

\renewcommand{\i}{\bm{i}}
\renewcommand{\j}{\bm{j}}
\renewcommand{\ij}{{\langle \bm{i}, \bm{j} \rangle}}
\newcommand{\ijprime}{{\langle \boldsymbol{i}^{\prime}, \boldsymbol{j}^{\prime} \rangle}}

\newcommand{\aj}{\hat{a}_{\bm{j}}}

\newcommand{\adj}{\hat{a}^\dagger_{\bm{j}}}

\newcommand{\nj}{\hat{n}_{\bm{j}}}

\newcommand{\Ztwo}{$\mathbb{Z}_2$}

\makeatletter
\def\maketitle{
\@author@finish
\title@column\titleblock@produce
\suppressfloats[t]}
\makeatother

\AtBeginDocument{%
    \newwrite\bibnotes
    \def\bibnotesext{Notes.bib}
    \immediate\openout\bibnotes=\jobname\bibnotesext
    \immediate\write\bibnotes{@CONTROL{REVTEX41Control}}
    \immediate\write\bibnotes{@CONTROL{%
    apsrev41Control,author="08",editor="1",pages="1",title="0",year="1"}}
     \if@filesw
     \immediate\write\@auxout{\string\citation{apsrev41Control}}%
    \fi
}%

\begin{document}

\title[Percolation as a confinement order parameter in Z2 lattice gauge theories]{Percolation as a confinement order parameter in $\mathbb{Z}_2$~lattice gauge theories}

\author{Simon M. Linsel}
\email{simon.linsel@physik.uni-muenchen.de}
\affiliation{Department of Physics and Arnold Sommerfeld Center for Theoretical Physics (ASC), Ludwig-Maximilians-Universit\"at M\"unchen, Theresienstr. 37, M\"unchen D-80333, Germany}
\affiliation{Munich Center for Quantum Science and Technology (MCQST), Schellingstr. 4, D-80799 M\"unchen, Germany}

\author{Annabelle Bohrdt}
\affiliation{Munich Center for Quantum Science and Technology (MCQST), Schellingstr. 4, D-80799 M\"unchen, Germany}
\affiliation{Institute of Theoretical Physics, University of Regensburg, D-93053, Germany}

\author{Lukas Homeier}
\affiliation{Department of Physics and Arnold Sommerfeld Center for Theoretical Physics (ASC), Ludwig-Maximilians-Universit\"at M\"unchen, Theresienstr. 37, M\"unchen D-80333, Germany}
\affiliation{Munich Center for Quantum Science and Technology (MCQST), Schellingstr. 4, D-80799 M\"unchen, Germany}

\author{Lode Pollet}
\affiliation{Department of Physics and Arnold Sommerfeld Center for Theoretical Physics (ASC), Ludwig-Maximilians-Universit\"at M\"unchen, Theresienstr. 37, M\"unchen D-80333, Germany}
\affiliation{Munich Center for Quantum Science and Technology (MCQST), Schellingstr. 4, D-80799 M\"unchen, Germany}

\author{Fabian Grusdt}
\email{fabian.grusdt@physik.uni-muenchen.de}
\affiliation{Department of Physics and Arnold Sommerfeld Center for Theoretical Physics (ASC), Ludwig-Maximilians-Universit\"at M\"unchen, Theresienstr. 37, M\"unchen D-80333, Germany}
\affiliation{Munich Center for Quantum Science and Technology (MCQST), Schellingstr. 4, D-80799 M\"unchen, Germany}

\date{\today}
\begin{abstract}
Lattice gauge theories (LGTs) were introduced in 1974 by Wilson to study quark confinement. These models have been shown to exhibit (de)confined phases, yet it remains challenging to define experimentally accessible order parameters. Here we propose percolation-inspired order parameters (POPs) to probe confinement of dynamical matter in \Ztwo{}~LGTs using electric field basis snapshots accessible to quantum simulators. We apply the POPs to study a classical \Ztwo{}~LGT and find a confining phase up to temperature $T=\infty$ in 2D (critical $T_c$, i.e. finite-$T$ phase transition, in 3D) for any non-zero density of \Ztwo{}~charges. Further, using quantum Monte Carlo we demonstrate that the POPs reproduce the square lattice Fradkin-Shenker phase diagram at $T=0$ and explore the phase diagram at $T>0$. The correlation length exponent coincides with the one of the 3D Ising universality class and we determine the POP critical exponent characterizing percolation. Our proposed POPs provide a geometric perspective of confinement and are directly accessible to snapshots obtained in quantum simulators, making them suitable as a probe for quantum spin liquids.
\end{abstract}
\maketitle


\textit{Introduction.---}Lattice gauge theories (LGTs) have been widely studied in the fields of high-energy~\cite{Wilson1974}, condensed matter~\cite{Wegner1971, FradkinSusskind1978, Kogut1979} and biophysics~\cite{LammertRokhsarToner1993}. In 1979, Fradkin and Shenker proved in their groundbreaking work~\cite{FradkinShenker1979} the existence of two phases in their model, where \Ztwo{}~charged particles are confined or deconfined, respectively. Ever since, researchers have found intimate connections of \Ztwo{}~LGTs to other physical effects, including topological order~\cite{Wen2007, Verresen2022}, quantum spin liquids~\cite{ReadSachdev1991, Sachdev2019} and even quantum information~\cite{Kitaev2003}. In the light of quantum simulation, LGTs with finite-dimensional local Hilbert spaces, e.g. \Ztwo{}~LGTs and quantum link models \cite{Wiese2013}, gain particular attention because of their experimental feasibility \cite{Halimeh2023}.

Despite experimental progress, it remains a challenging problem to define order parameters for deconfined (e.g. topological) phases that are accessible to both numerical simulations and cold-atom experiments. Wegner-Wilson loops (WWL) \cite{Wegner1971, Wilson1974} are non-local order parameters allowing to probe (de)confinement in pure gauge theories, i.e. without matter. They measure the fluctuation of the magnetic field and feature an area (perimeter) law in the (de)confined phase. However, they are not suitable for \Ztwo{}~LGTs with matter where they follow a perimeter law regardless of the phase \cite{FradkinShenker1979}. The Fredenhagen-Marcu order parameter \cite{Fredenhagen1983, Fredenhagen1986, Fredenhagen1988} solves this problem by relating a ``full''- to a ``half'' WWL, measuring the response of the system when spatially separating two 
matter particles \cite{Gregor2011}. Despite being a ratio of two small numbers, the Fredenhagen-Marcu order parameter was used experimentally for systems with a strong Rydberg blockade \cite{Semeghini2021, Verresen2021}.

Much physical intuition of confinement of matter comes from electric field lines connecting gauge charges, creating a linear confining potential \cite{Zhang2023}. In the electric field basis -- accessible to state-of-the-art quantum simulators \cite{Zohar2017, Barbiero2019, Schweizer2019_2, Homeier2021, Homeier2022, Mildenberger2022, Halimeh2023} -- a picture of fluctuating electric fields is appealing, analogous to fluctuating magnetic fields in WWL: if the electric field lines are fluctuating so strongly that one cannot distinguish whether two charges are connected or not, the picture of a confining electric string breaks down. As will be shown in this Letter, this physical intuition can be formalized with the help of percolation theory, which has been used extensively to study phase transitions geometrically~\cite{Essam1980}.

Site percolation theory was first introduced in works by Flory and Stockmayer in the 1940s~\cite{Flory1941, Stockmayer1944}, where they studied the gel-point and the cross-linking of polymers. Nearly two decades later, bond percolation theory was introduced by Broadbent and Hammersley~\cite{BroadbentHammersley1957}. They studied the random flow of a fluid through a medium and introduced the so-called \textit{Bernoulli percolation}. A prime example of the use of percolation theory in physics is the random cluster (RC) model, which was introduced by Fortuin and Kasteleyn~\cite{Kasteleyn1969, Fortuin1972_1, Fortuin1972_2, Fortuin1972_3}. The RC model was successfully used to study many physical systems, e.g. the Potts model~\cite{Grimmett2006, Baxter1985}. More recently, percolation theory was applied in the context of quantum monopole motion in spin ice \cite{Stern2021}, quantum error correction \cite{Hastings2014} and gauge theories \cite{Gliozzi2005}.

In this Letter, we show that percolation can also serve as a non-local order parameter to probe confinement. In the confined phase, analogous to quarks in a confining potential, two charges are being held together by \Ztwo{} electric strings and form a \Ztwo{}-neutral meson \cite{Homeier2022}. In the deconfined phase, in our picture matter particles move essentially independently through the system and the \Ztwo{} electric strings form a global cluster spanning over the entire lattice, see Fig.~\ref{fig1}. Charges attached to this cluster are no longer constrained to \Ztwo{} neutral hadrons. We study a classical and a quantum model using classical (quantum) Monte Carlo (MC/QMC). We show that the phase boundary in the celebrated Fradkin-Shenker model (extended toric code)~\cite{FradkinShenker1979} can be reproduced by percolation-inspired order parameters (POPs), paving the way for the application of POPs in related models.

\begin{figure}[t]
\includegraphics[width=0.45\textwidth]{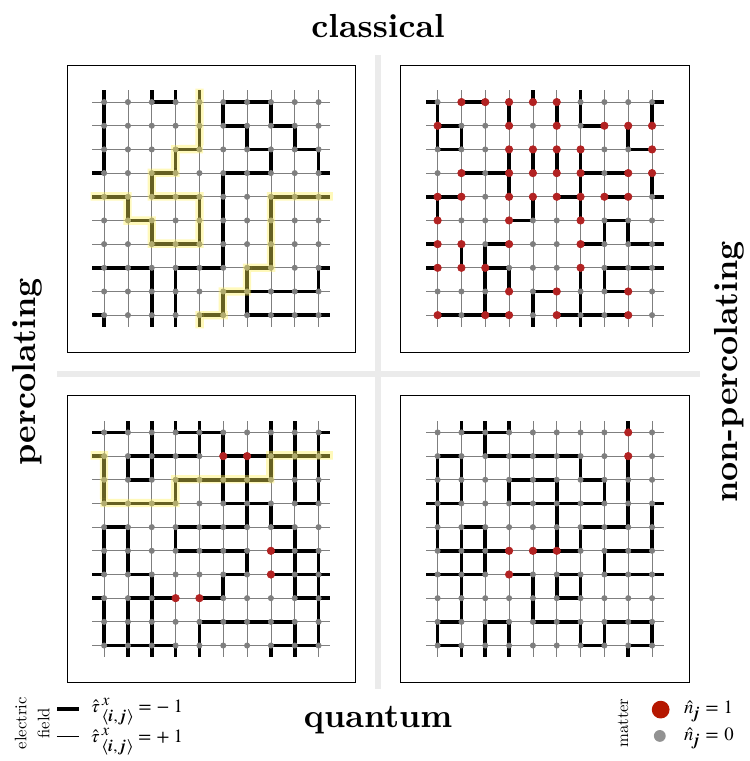}
\caption{\textbf{Classical and quantum snapshots of \mbox{(non-)} percolating strings.} We show periodic classical/quantum snapshots of (non-)percolating strings obtained from MC simulations. Here, percolation refers to a cluster of electric fields winding around the periodic system and connecting to itself in at least one dimension, see \cite{Supplements} Sec.~\ref{app:classical_mc_sampling}. The \mbox{(non)}\-percolating phase is associated with a deconfined (confined) phase. \textbf{Top:} The classical snapshots are obtained from Hamiltonian~(\ref{eq:can_cl_ham}). The percolating snapshot is captured at high temperature for a system without matter excitations. Introducing matter prevents the formation of a percolating string, as can be observed in the confined snapshot. \textbf{Bottom:} The ground state quantum snapshots are obtained from equal imaginary time slices of Hamiltonian~(\ref{eq:eTC}), inside (percolating) and outside (non-percolating) the topological phase. Here, a non-zero matter density does not necessarily prohibit percolation.}
\label{fig1}
\end{figure}


\textit{Classical \Ztwo{}~LGT.---} The complex nature of percolation in \Ztwo{}~LGTs emerges from the local \Ztwo{} symmetry generator 
\begin{align}
    \Gj = (-1)^{\nj} \prod_{l \in +_{\j}}\hat{\tau}_{l}^x, \label{eq:Goperator}
\end{align}
where~$\nj=\adj\aj$ is the number operator for (hard-core, Higgs) matter on site~$\j$ and the Pauli matrix~$\hat{\tau}^x_l$ defines the electric field on a link $l$ adjacent to site~$\j$. This leads to an extensive set of conserved quantities $\{g_{\bm{j}}\}$ which fulfill Gauss's law $\Gj \ket{\psi} = g_{\bm{j}}\ket{\psi}$. The choice of $\{g_{\bm{j}}\}$ defines a gauge sector. In the following, we only consider $g_{\bm{j}}=1$, i.e. no background charges.

\begin{figure}[t]
\includegraphics[width=0.45\textwidth]{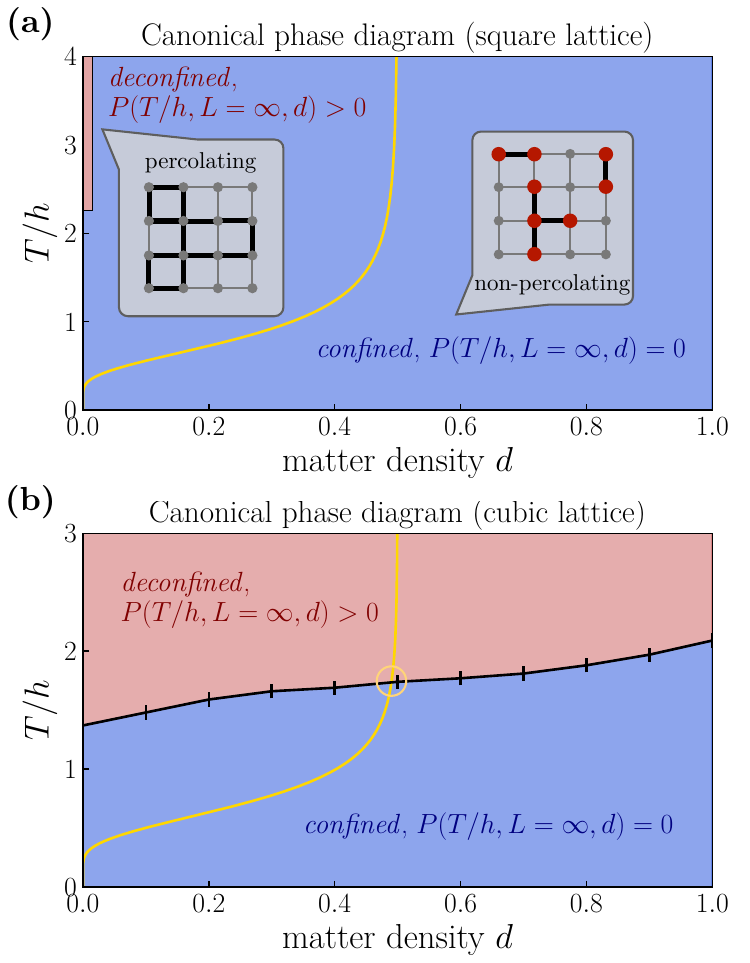}
\caption{\textbf{Canonical thermal deconfinement phase diagrams at non-zero matter density.} We show two temperature-density phase diagrams of the classical Hamiltonian~(\ref{eq:can_cl_ham}) for the square \textbf{(a)} and the cubic lattice \textbf{(b)}. In the confined phase (blue), the percolation strength $P(T/h, L=\infty, d)$ vanishes while strings percolate $P(T/h, L=\infty, d) > 0$ in the deconfined phase (red). The yellow lines are the exactly solvable $\mu$=0-case of Hamiltonian~(\ref{eq:gc_cl_ham}), see Eq.~(\ref{eq:p}, \ref{eq:d}). In panel~\textbf{(a)}, we find a thermal deconfinement transition only at zero matter density. The critical temperature $(T/h)_c = 2.27(1)$, extracted via a finite-size scaling analysis, is identical to the Ising critical temperature. At non-zero matter density, the system is confining for all temperatures. In panel~\textbf{(b)}, we present results for the cubic lattice. In contrast to the square lattice, a finite-temperature thermal deconfinement transition exists for arbitrary matter density. At the $\mu$=0-line, the phase transition to the deconfined phase (yellow circle) matches the Bernoulli percolation threshold $p_{\mathrm{c,cu}}=0.247(5)$ \cite{Sykes1964} (note that we show $d$, not $p$).}
\label{fig2}
\end{figure}

To demonstrate the viability of the POPs, we first perform MC simulations of the \Ztwo{}-symmetric classical Hamiltonian
\begin{align} \label{eq:can_cl_ham}
    \hat{\mathcal{H}}_{\mathrm{can}} = - h \sum_{l} \hat{\tau}_{l}^x
\end{align}
in the canonical regime, i.e. we fix the number of matter particles $N=\sum_{\j} \hat{n}_{\j}$. We assume $h>0$. 

Due to Gauss's law, each matter particle has to be connected to a string $\Sigma$ of electric fields where $\hat{\tau}^x_l = -1 \; \forall l \in \Sigma$. This string costs an energy $2h\ell$, where $\ell$ is the length of the string. At low temperature $T$ and low matter density, this results in matter particles forming mesonic bound states, where matter particles on neighboring sites are connected by a string of length one \cite{Homeier2022}. At higher temperatures, there is a competition between the entropy $S(\ell)$ and the energy $E(\ell)$ of strings (in linear approximation). When $T S > E$, a global string network forms that winds around the system for periodic boundaries while matter particles become free \Ztwo{}~charges \cite{Hahn2022}. We describe this as a percolation transition of the strings from a non-percolating bound mesonic regime to a percolating regime with incoherent but free charges.

To probe this transition, we define two POPs. \textit{Percolation probability} is the probability that a cluster percolates, i.e. winds around the system and connects to itself, see \cite{Supplements} Sec.~\ref{app:can_mc_sampling}. In the literature, this concept is sometimes also referred to as the wrapping probability \cite{Feng2008} or the winding number \cite{Banerjee2022}. \textit{Percolation strength} extends the percolation probability to include information about the size of the percolating cluster. When the system is not percolating, the percolation strength is zero. When the system is percolating, the percolation strength is defined as the number of strings in the largest string cluster divided by the total number of links \cite{Homeier2022}. We probe the transition on a periodic square lattice with system sizes $L^2 = 20^2, 30^2, 40^2, 50^2$. In the following we distinguish the cases for (i) zero matter density $d=0$ and (ii) non-zero matter density $d=N/L^2 > 0$.

(i) Hamiltonian~(\ref{eq:can_cl_ham}) can be mapped to the classical 2D Ising model on the dual lattice \cite{Wegner1971, Fradkin1978}, where the confined phase corresponds to the ferromagnetic ordered phase. The POPs can be viewed as a \textit{dis}order parameter detecting domain walls in the ferromagnet, see \cite{Supplements} Sec.~\ref{app:ising_mapping}. Via a finite-size scaling analysis, we confirm that the percolation transition we find takes place exactly at the Ising critical temperature $(T/h)_c \vert_{\mathrm{Ising}} = 2 / \log(1+\sqrt{2}) \approx 2.27$ \cite{Onsager1944} (see Fig.~\ref{fig2}a). For the correlation length critical exponents, we find $\nu = 1.04(10)$ (2D Ising universality). Hence, one divergent length scale for percolation and Ising spins fixes $\nu$. This is consistent with previous studies, where the exponent $\nu$ of the confinement length \cite{Trebst2007} and wrapping probabilities \cite{Huang2020} was found to be of Ising universality. For the percolation strength critical exponent, we find $\beta = 0.58(10)$. We are unaware of a direct relation of the percolation strength critical exponent $\beta$ to the 2D Ising critical exponent. The percolation probability can also serve as an order parameter but it is impossible to extract a corresponding $\beta$ since the percolation probability jumps from zero to one in the thermodynamic limit (TDL).

(ii) We numerically find no thermal deconfinement transition, i.e. the presence of matter prevents the formation of a percolating string cluster and is thus always confined in the TDL, see \cite{Supplements} Sec.~\ref{app:classical_finite_matter_density}. To unravel this behavior analytically, we study the grand-canonical version of Hamiltonian~(\ref{eq:can_cl_ham}),
\begin{align} \label{eq:gc_cl_ham}
    \hat{\mathcal{H}}_{\mathrm{gc}} &= - h \sum_{l} \hat{\tau}_{l}^x - \mu \sum_{\j} \hat{n}_{\j} \nonumber \\
    &= - h \sum_{l} \hat{\tau}_{l}^x - \mu \sum_{\j} \frac{1}{2} \Bigl( 1- \prod_{l \in +_{\j}}\hat{\tau}_{l}^x \Bigr),
\end{align}
where we used the Gauss law constraint $\G_{\j} = +1$, Eq.~(\ref{eq:Goperator}), to express $\hat{n}_{\j}$ by $\hat{\tau}^x_l$ in the second line. The chemical potential $\mu$ serves as a Lagrange multiplier for the matter density. Eq.~(\ref{eq:gc_cl_ham}) is a generalized Ising model with four-spin interactions $\propto \mu$. 

\begin{figure}[t]
\includegraphics[width=0.45\textwidth]{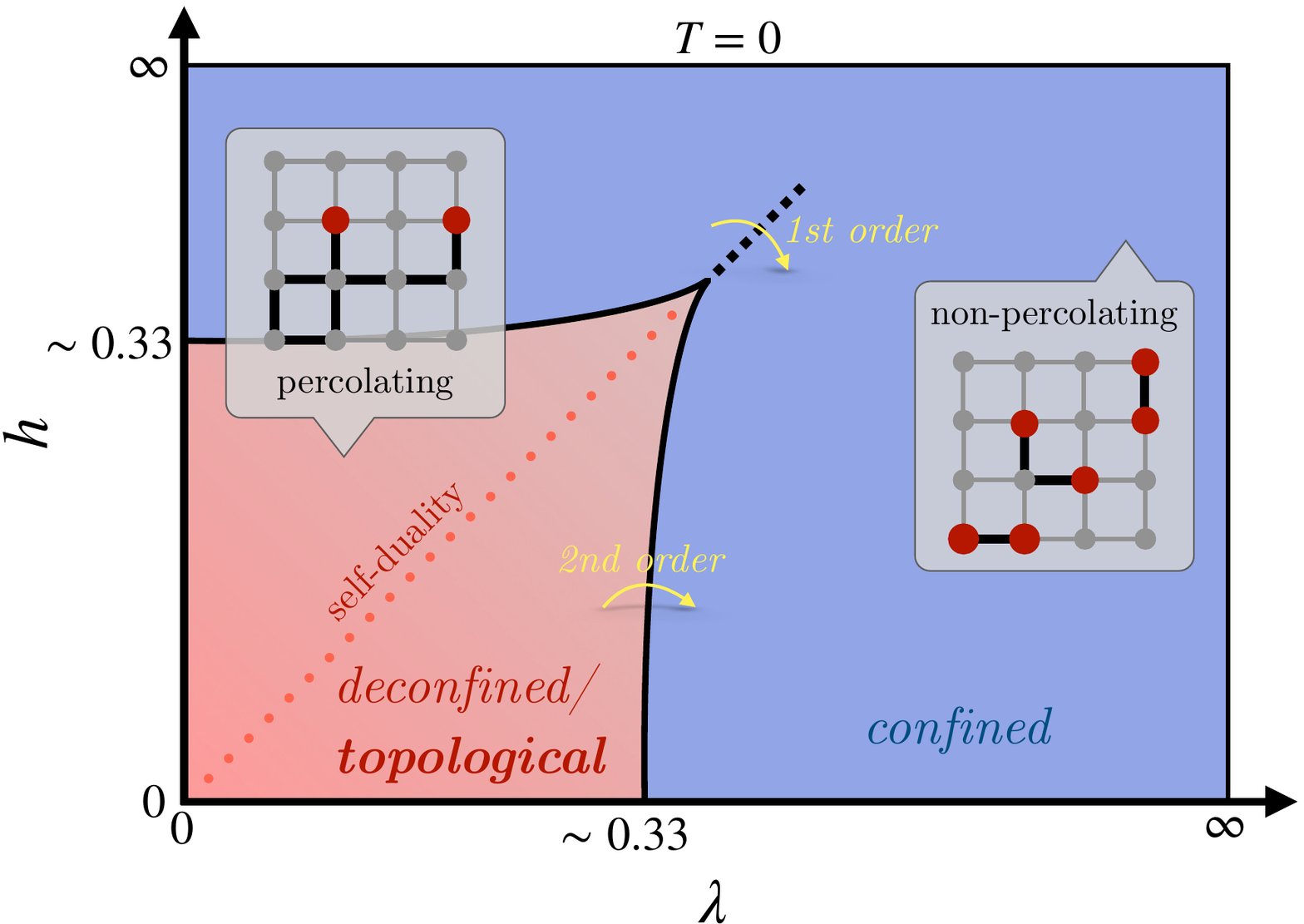}
\caption{\textbf{Fradkin-Shenker phase diagram.} We show that percolation identifies the same phase boundaries for the Fradkin-Shenker phase diagram of the extended toric code ($\mu=J=1$) \cite{FradkinShenker1979, Wu2012} on the square lattice, see Hamiltonian~(\ref{eq:eTC}): in the deconfined phase (red) electric strings percolate while the percolation strength vanishes in the confined phase (blue). The confinement transition is continuous (solid lines) while there is a first-order line (dashed line) extending into the confined phase \cite{Wu2012}.}
\label{fig3}
\end{figure}

At $\mu = 0$, the electric fields are completely independent resembling standard Bernoulli bond percolation. The corresponding probability $p$ for a given bond to host a string (i.e. $\hat{\tau}^x_l=-1$) is 
\begin{align} \label{eq:p}
    p = e^{-\beta h} / [ 2 \cosh(\beta h) ]
\end{align}
with $\beta = 1/T$, from independent thermal distributions at each bond. Hence, $0 \leq p < 1/2$ for $\mu = 0$, which we find to correspond to average matter densities 
\begin{align} \label{eq:d}
    d=\frac{1}{2} \bigl[ 1-(2p-1)^z \bigr]
\end{align}
for a lattice with even coordination number $z$. Importantly, all values of $p$ remain below the Bernoulli percolation threshold $p_{\mathrm{c,sq}}=0.5$ on the square lattice \cite{Sykes1964, Kesten1980}. Hence, for $\mu = 0$ (corresponding to $d=1/2$ for $T/h \to \infty$) we analytically proved that the presence of matter prohibits percolation, i.e. there is no finite-$T$ percolation phase transition, see Fig.~\ref{fig2}a. Notably, the percolation threshold is reached when $T/h \to \infty$, indicating a deconfined, critically percolating state at infinite temperature.

Conversely, the percolation threshold for the cubic lattice is $p_{\mathrm{c,cu}}=0.247(5)$ \cite{Sykes1964}, implying that a thermal deconfinement phase transition at finite $T$ can exist at any matter density $d$ since $p \to 1/2$ for $T \to \infty$. We simulate the periodic cubic lattice with system sizes $L^3 = 10^3, 12^3, 14^3$ and show the phase diagram in Fig.~\ref{fig2}b. As expected, and in contrast to the square lattice, the deconfined phase persists for arbitrary matter density and sufficiently high temperatures. 



\textit{Quantum \Ztwo{}~LGT: extended toric code.---}We start with a generalization of Hamiltonian~(\ref{eq:gc_cl_ham}) by adding magnetic fluctuations $\propto J$ and dynamical matter $\propto \lambda$:
\begin{align} \label{eq:gen_gc}
    \hat{\mathcal{H}}_{\mathrm{qLGT}} = &- \mu \sum_{\j} \nj \; - h \sum_{l} \hat{\tau}_l^x \; - J \sum_{\square} \prod_{l \in \square} \hat{\tau}_l^z \nonumber \\
    &- \lambda \sum_{\ij} (\hat{a}_{\i}^\dagger + \hat{a}_{\i}) \, \hat{\tau}_l^z \, (\hat{a}_{\j}^\dagger + \hat{a}_{\j}).
\end{align}
This quantum Hamiltonian fulfills $[\hat{\mathcal{H}}_{\mathrm{qLGT}}, \Gj] = 0$ and is thus a \Ztwo{}~LGT. To study this Hamiltonian with QMC, we restrict ourselves to the gauge sector \mbox{$g_{\j} = + 1$}. We integrate out the matter fields and write 
\begin{align}
    \nj = \frac{1}{2} \biggl( 1 - \prod_{l \in +_{\j}}\hat{\tau}_{l}^x \biggr). \label{eq:simplifiedGoperator}
\end{align}
I.e. the configuration $\{\hat{\tau}_l^x\}$ uniquely determines the state of the hard-core matter. Using Eq.~(\ref{eq:simplifiedGoperator}) in Hamiltonian~(\ref{eq:gen_gc}) yields
\begin{align} \label{eq:eTC}
    \hat{\mathcal{H}}_{\mathrm{eTC}} = &- \mu \sum_+ \prod_{l \in +} \hat{\tau}_l^x \; - J \sum_{\square} \prod_{l \in \square} \hat{\tau}_l^z \nonumber \\
    &- h \sum_{l} \hat{\tau}_l^x \; - \lambda \sum_{l} \hat{\tau}_l^z,
\end{align}
exactly and \textit{without} local constraints. This is the extended toric code where in LGT language, $h$ acts as a confining potential and $\lambda$ can be viewed as a gauge-breaking perturbation in the pure gauge theory. This model was originally studied by Fradkin and Shenker \cite{FradkinShenker1979}, featuring a $T=0$ phase diagram with a deconfined topological phase for small fields and a confined phase for large fields $h,\lambda$ \cite{FradkinShenker1979, Wu2012}, see Fig.~\ref{fig3}. Fradkin and Shenker \cite{FradkinShenker1979} proved that the confined ($h \gg 1$) and Higgs phase ($\lambda \gg 1$) are adiabatically connected and thus identical. In the following, we fix $\mu = J = 1$.

In order to study percolation in the quantum \Ztwo{}~LGT, we adapt the continuous-time QMC algorithm from Wu~\textit{et al.}~\cite{Wu2012} to the $\hat{\tau}^x$-basis to extract snapshots of strings and set $T = 1/L$ to gain insights into the ground state phase diagram. Our analysis of such data leads us to the main result of this Letter: we conjecture that confinement in the quantum \Ztwo{}~LGT with dynamical matter can be directly probed through percolation of \Ztwo{} electric strings.

The extended toric code features a self-duality ($h \leftrightarrow \lambda$) and the pure gauge model ($\lambda=0$) can be mapped to the 2D transverse-field Ising model on the dual lattice \cite{Wegner1971, Kogut1979, Supplements} where the quantum critical point is in the 3D Ising universality class. Above the self-duality line, i.e. $h>\lambda$, we find that the POPs reproduce the well-known phase transition from a deconfined (percolating) regime to a confined (non-percolating) phase for the entire phase boundary. We perform a finite-size scaling analysis to extract the critical temperature and the critical exponents. We illustrate a crossing-point analysis of the percolation strength Binder cumulant at $\lambda=0.3$ in Fig.~\ref{fig4}a. The correlation length critical exponent $\nu=0.65(27)$ matches the 3D Ising universality class. This is akin to the phenomenology we found for the classical \Ztwo{}~LGT above. For the percolation strength critical exponent we find $\beta=0.05(4)$. We were unable to provide a good estimate of the Fredenhagen-Marcu order parameter in the confined phase for $h > \lambda$ due to high noise in the data.

\begin{figure}[t]
\includegraphics[width=0.45\textwidth]{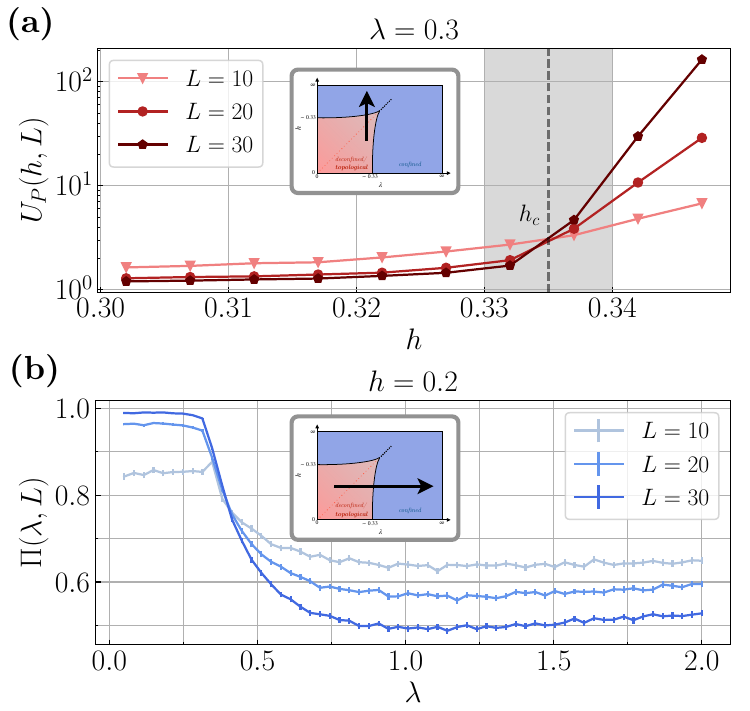}
\caption{\textbf{QMC results of the extended toric code.} We show results for the POPs in the square lattice toric code Hamiltonian~(\ref{eq:eTC}) at $T=1/L$. In panel~\textbf{(a)}, we present a crossing-point analysis of the percolation strength Binder cumulant $U_P$ above the self-duality line at $\lambda=0.3$, i.e. at non-zero charge density. The critical electric field $h_c=0.335(2)$ is in very good agreement with previous studies of the extended toric code \cite{Wu2012}. In panel~\textbf{(b)}, we present the percolation probability $\Pi(\lambda)$ for different system sizes at $h=0.2$. At large couplings $\lambda$, we observe a clear decrease in $\Pi$ with higher system sizes, and we conjecture that in the confined phase $\Pi$ approaches zero in the TDL.}
\label{fig4}
\end{figure}

Below the self-duality line, i.e. $h<\lambda$, the $\hat{\tau}^x$-basis is numerically less efficient in sampling snapshots than the $\hat{\tau}^z$-basis (since spins tend to be more aligned in the $\hat{\tau}^z$-basis), however, we still find a percolation transition in the $\hat{\tau}^x$-basis in this regime. From self-duality, we know that the physical phase transition is at $\lambda_c \approx 0.335$. For $h=0.2$, we find Binder cumulant crossing points at $\lambda \approx 0.4$ that shift toward lower $\lambda$ for increasing system size. We argue that this behavior is a finite-size effect and can be traced back to the finite-size gap, see \cite{Supplements} Sec.~\ref{app:qmc}. This view is supported by the decrease of the percolation probability we observe with increasing system size, see Fig.~\ref{fig4}b, although it does not yet extrapolate to zero for the simulated system sizes up to $L^2 = 30^2$. For Bernoulli percolation, Kolmogorov’s zero-one-law \cite{Lyons1990} restricts the percolation probability to either zero or one in the TDL. Even though the percolation of electric strings is correlated, we conjecture that the percolation probability in the confined phase approaches zero in the TDL. We are also able to probe the transition with the Fredenhagen-Marcu operator, see \cite{Supplements} Sec.~\ref{app:qmc}.

At $h=\lambda=0.2$ (deconfined, topological ground state), we find a percolating phase at low but finite temperatures and a non-percolating phase at higher temperatures. We trace this back to the bulk gap, where due to exponential suppression of thermal matter excitations the topological ground state stays robust against a non-zero temperature in finite-size systems, see \cite{Supplements} Sec.~\ref{app:qmc}. At temperatures well above the bulk gap, we observe a clear decrease in the percolation strength with higher system sizes. We conjecture that the percolation probability approaches zero in the TDL for any $T>0$. We were unable to provide a good estimate of the Fredenhagen-Marcu order parameter for finite temperatures due to high noise in the data. In the literature, confinement at finite temperature is also probed using Polyakov loops \cite{Banerjee2024} or by explicitly violating the gauge symmetry \cite{Banerjee2013}.

Our results for the quantum \Ztwo{}~LGT can be interpreted by drawing connections with the classical \Ztwo{}~LGT studied above. On the pure gauge axis, without matter, both support finite-$T$ percolation transitions which can be related to standard WWL. In the square lattice classical theory we find a complete breakdown of percolation at any non-vanishing matter density and for any $T$: Likewise, percolation breaks down in the Higgs phase (large $\lambda$) where \Ztwo{} charges accumulate and condense. In contrast, the $T=0$ deconfined phase in the quantum theory sustains percolation, because \Ztwo{} charges are gapped and only appear virtually as correlated pairs in the ground state. This picture changes once again when thermal fluctuations are included: these lead to a non-zero density of thermally activated but free \Ztwo{}-charged excitations at any $T>0$, where the classical theory explains the breakdown of string percolation.

\textit{Discussion and outlook.---}We have introduced new POPs to probe the confinement of matter excitations in \Ztwo{}~LGTs. We have applied it to study a classical model where we demonstrated the substantial influence of matter and the lattice geometry on the existence of a percolating, i.e. deconfined, phase. Further, using snapshots generated from continuous-time QMC, we demonstrated that the phase boundaries of the square lattice Fradkin-Shenker model, one of the textbook examples featuring confinement and topological order, can be reproduced by our POPs. The correlation length critical exponent falls into the 3D Ising universality class. We also demonstrated the potential of the POPs at finite temperature. Another task, left for future research, will be to check for a first-order percolation transition across the self-duality line at the tip of the deconfined phase.

While we only discuss \Ztwo{}~LGTs on the square and the cubic lattice, we also envision future applications of the POPs in other LGTs, where different interactions, lattice geometries or higher symmetries can be realized. Percolation was already used in the context of quantum error correcting codes \cite{Botzung2023}. Since percolation can be directly measured from snapshots in the electric field basis, the order parameter is readily accessible to state-of-the-art quantum simulators \cite{Zohar2017, Barbiero2019, Schweizer2019_2, Homeier2021, Homeier2022, Mildenberger2022, Halimeh2023}, and extends the geometric perspective on confinement beyond one-dimensional systems \cite{Kebric2021, Kebric2023}.\medskip

We thank M. Aidelsburger, D. Banerjee, G. Dünnweber, M. Kebri\v c, F. Palm, G. Semeghini, T. Wang and H. Zhao for fruitful discussions.
This research was funded by the European Research Council (ERC) under the European Union’s Horizon 2020 research and innovation program (Grant Agreement no 948141) — ERC Starting Grant SimUcQuam, and by the Deutsche Forschungsgemeinschaft (DFG, German Research Foundation) under Germany's Excellence Strategy -- EXC-2111 -- 390814868 and via Research Unit FOR 2414 under project number 277974659. L.H. acknowledges support from the Studienstiftung des deutschen Volkes.

\newpage~
\newpage~

\setcounter{equation}{0}
\setcounter{figure}{0}
\setcounter{table}{0}
\setcounter{section}{0}
\setcounter{page}{1}
\renewcommand{\theequation}{S\arabic{equation}}
\renewcommand{\thefigure}{S\arabic{figure}}
\renewcommand{\thetable}{S\Roman{table}}

\title{Supplemental Material: Percolation as a confinement order parameter in $\mathbb{Z}_2$~lattice gauge theories}
\maketitle

\onecolumngrid

\section{Classical Monte Carlo sampling}
\label{app:classical_mc_sampling}

Here we present the numerical details of the classical MC sampling and related algorithms used to obtain snapshots, from which the observables for characterizing percolation are evaluated.

\subsection{Canonical Monte Carlo Sampling} \label{app:can_mc_sampling}
We simulate the classical Hamiltonian~(\ref{eq:can_cl_ham})
\begin{align} 
    \hat{\mathcal{H}}_{\mathrm{can}} = - h \sum_{l} \hat{\tau}^x_l \nonumber
\end{align}
on the square and the cubic lattice. Since we consider a canonical ensemble, we fix the number of matter particles $N=\sum_{\j} \hat{n}_{\j}$, with $(-1)^{\hat{n}_{\j}} = \prod_{l \in +_{\j}} \hat{\tau}^x_l$, in the system and all chemical potential terms are irrelevant. The size of the lattice determines the matter density resolution that can be achieved.

The MC simulations are implemented in C++ using the Boost C++ libraries. The lattice is represented by a graph. To generate (pseudo-)random numbers, we use the Mersenne Twister \cite{Matsumoto1998}. For I/O operations, multiprocessing and postprocessing, we use Python (NumPy, SciPy, Python multiprocessing, Matplotlib, h5py).

At the beginning of each simulation, we initialize the system in a state with the desired number of matter particles that fulfills Gauss's law. Matter particles can only be put into the system in pairs since Gauss's law would otherwise not be fulfilled. This is easily achieved by initializing matter in pairs connected by a string of length one (``dimers''). We then start the thermalization by repeatedly applying MC updates to this state.

\begin{figure}[t]
    \centering
    \includegraphics[width = 0.95 \textwidth]{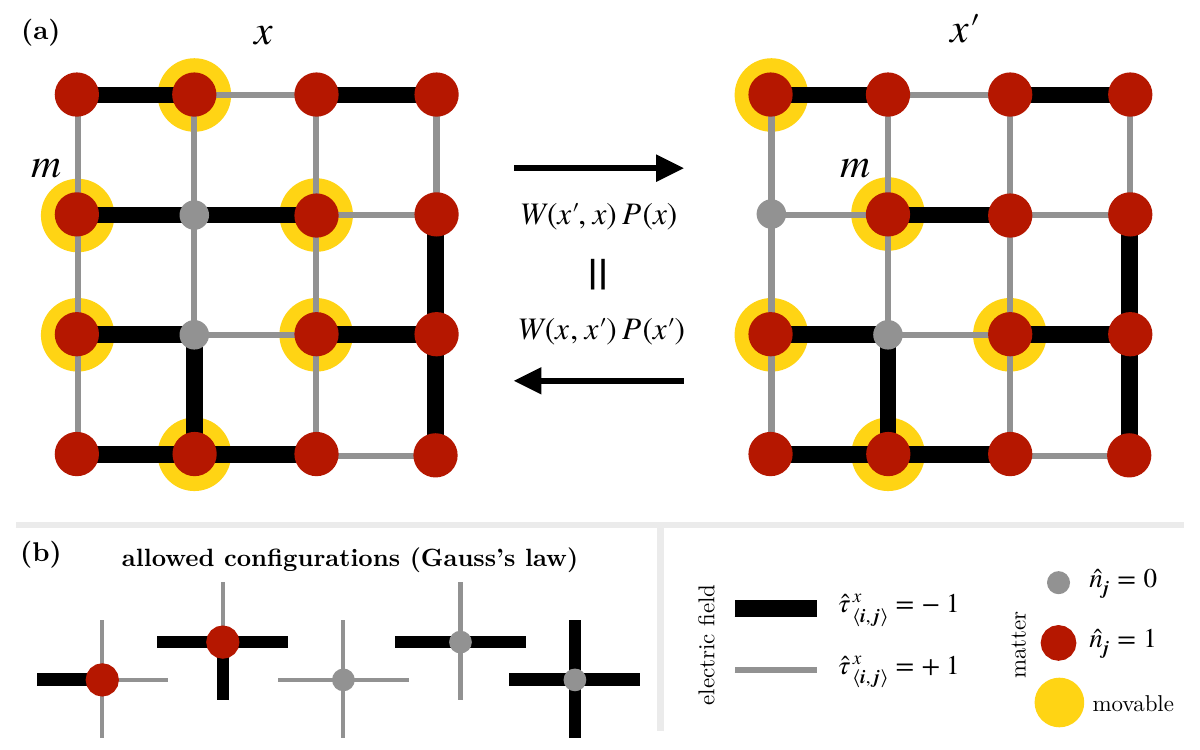}
    \caption[Ergodicity condition in canonical sampling]{\textbf{Ergodicity condition in canonical sampling.} In panel \textbf{(a)} we plot exemplary Metropolis-Hastings updates of a matter excitation $m$ moving between states $x$ and $x^\prime$. The Boltzmann distribution $P(x)$ and the transition rate $W(x^\prime ,\, x)$ from $x$ to $x^\prime$ must fulfill the detailed balance condition $W(x^\prime, \, x) P(x) = W(x, \, x^\prime) P(x^\prime)$ to represent a physical state in equilibrium. Hence, the number of unoccupied NN of $m$ and the total number of movable matter particles (marked in yellow) have to be included in the acceptance probability (see Eq.~(\ref{eq:can_samp_acceptance})). In panel \textbf{(b)}, we illustrate the constraints introduced by Gauss's law. Adapted from \cite{Homeier2022}.}
    \label{fig:can_ergodicity}
\end{figure}

A naive way to perform an MC update would be to choose a random matter particle and try to move it to a random nearest neighbor (NN) while flipping the electric field between the old site and the target site if the move is accepted to account for Gauss's law. If the target lattice site is already occupied, the move would be rejected because of the hard-core constraint. Because choosing any matter particle and any NN is equally likely (disregarding boundary effects), detailed balance would be trivially fulfilled. However, this approach becomes inefficient for large matter density, as a random matter particle is unlikely to be movable to \textit{any} NN (see Fig.~\ref{fig:can_ergodicity}a). Moreover, even if a matter particle is movable, it may be only movable to one NN which makes it highly likely that the move is rejected because of the randomness of the NNs in the trial probability distribution (TPD), see Eq.~(\ref{eq:TPD}). For this reason, the naive approach is not suitable for studying systems with a high matter density.

A better approach takes into account 1) only movable matter particles (i.e. matter with at least one unoccupied NN) and 2) only directions where no other matter particle is sitting. Hence, a move is never rejected because of the hard-core constraint (although it may be rejected because of the acceptance probability). This approach is orders of magnitude more efficient than the naive way, but it comes at the cost of a more complicated procedure to fulfill detailed balance.

Firstly, the number of movable matter particles can change after a move (illustrated in Fig.~\ref{fig:can_ergodicity}a), which makes the TPD asymmetric. Secondly, the number of empty NNs of one given matter particle can change when it moves (illustrated in Fig.~\ref{fig:can_ergodicity}a), which again makes the TPD asymmetric. Hence, we need the Metropolis-Hastings algorithm and the simpler Metropolis algorithm is \textit{not} suited. 

Let us suppose we propose a move of a matter particle $h$ from state $x$ to $x^\prime$. The TPD of this move is given by
\begin{align} \label{eq:TPD}
    T(x^\prime \, | \, x) = \frac{1}{\#(\text{movable matter at} \, x)} \times \frac{1}{\#(\text{NNs of matter} \, h \, \text{at} \, x)} 
\end{align}
while the TPD for a move from $x^\prime$ to $x$ is given by 
\begin{align}
    T(x \, | \, x^\prime) = \frac{1}{\#(\text{movable matter at} \, x^\prime)} \times \frac{1}{\#(\text{NNs of matter} \, h \, \text{at} \, x^\prime)}. \nonumber
\end{align}
Hence, the Metropolis-Hastings acceptance probability reads 
\begin{align} \label{eq:can_samp_acceptance}
    A(x^\prime, x) &= \min \biggl(1, \frac{e^{-\beta E(x^\prime)}}{e^{-\beta E(x)}} \frac{T(x \,|\, x^\prime)}{T(x^\prime \,|\, x)} \biggr) \nonumber \\
    &= \min \biggl( 1, \frac{e^{-\beta E(x^\prime) }}{e^{-\beta E(x)}} \times \frac{\#(\text{NNs of matter} \, h \, \text{at} \, x)}{\#(\text{NNs of matter} \, h \, \text{at} \, x^\prime)} \times \frac{\#(\text{movable matter at} \, x)}{\#(\text{movable matter at} \, x^\prime)} \biggr). 
\end{align}
For efficiency reasons, we store the ``movability'' of every matter particle and update this information after every successful move update. Only matter particles in the neighborhood of the move update must get a movability check, otherwise, there would be no point in storing the information altogether.

\begin{figure}[t]
    \centering
    \includegraphics[width = 0.95 \textwidth]{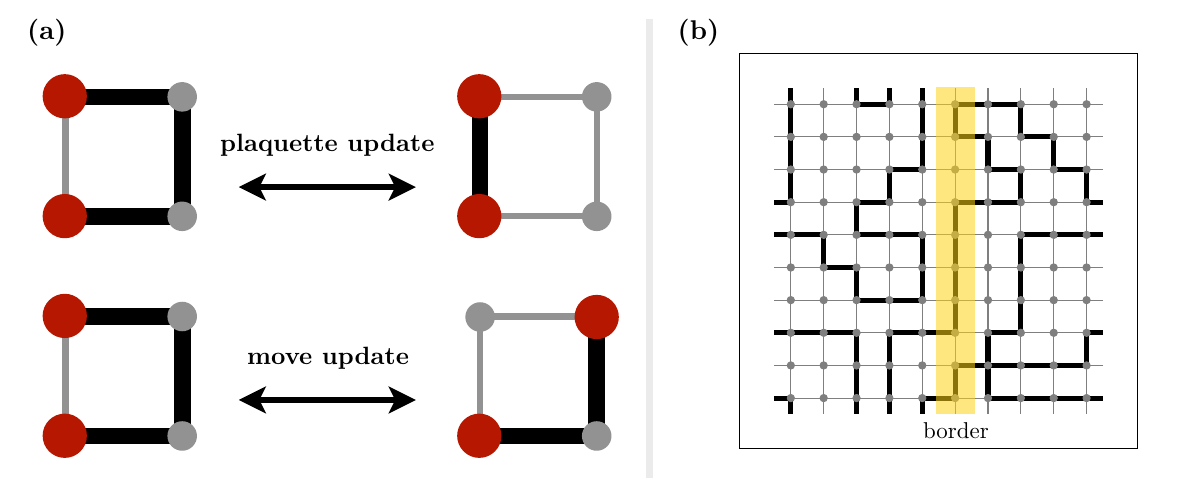}
    \caption[MC updates and percolation detection]{\textbf{MC updates and percolation detection.} In panel \textbf{(a)}, we show the update procedures for the Metropolis-Hastings sampling on the square lattice. A plaquette update flips all electric fields in a plaquette. In a move update, a matter particle moves to an unoccupied lattice site and flips the electric field along its way. All updates conserve the number of matter particles and Gauss’s law. In panel \textbf{(b)} we sketch the algorithm to detect percolation in $x$-direction (horizontal). The algorithm tries to find a path that winds around the lattice while only traversing strings. In this example, the strings percolate. When crossing the border, sites get assigned a higher/lower winding number, see Alg.~\ref{alg:percolation}. Adapted from \cite{Homeier2022}.}
    \label{fig:mc_updates}
\end{figure}

If we only used the move update, we would face two problems: 1) It does not work for the cases where we have no matter or every lattice site is occupied by a matter particle (density $d=1$), since in both cases no matter particle is movable, and 2) it can be slow for high temperatures since only one string is created (or removed) for every successful MC update. For this reason, we introduce plaquette updates (see Fig.~\ref{fig:mc_updates}a), which flip all electric fields in an elementary plaquette. Plaquette updates can create strings up to the length of the plaquette and thus speed up the simulations for high temperatures, where higher excited states with more strings are more likely. In the TPD, every elementary plaquette is equally likely to be tried. Hence, we have a symmetric TPD and Metropolis sampling is sufficient. In every MC update, there is a 50\% chance to propose either a plaquette update or a move update (except at zero or maximal matter density, where we only use plaquette updates). 

\begin{figure}[t]
\includegraphics[width=\textwidth]{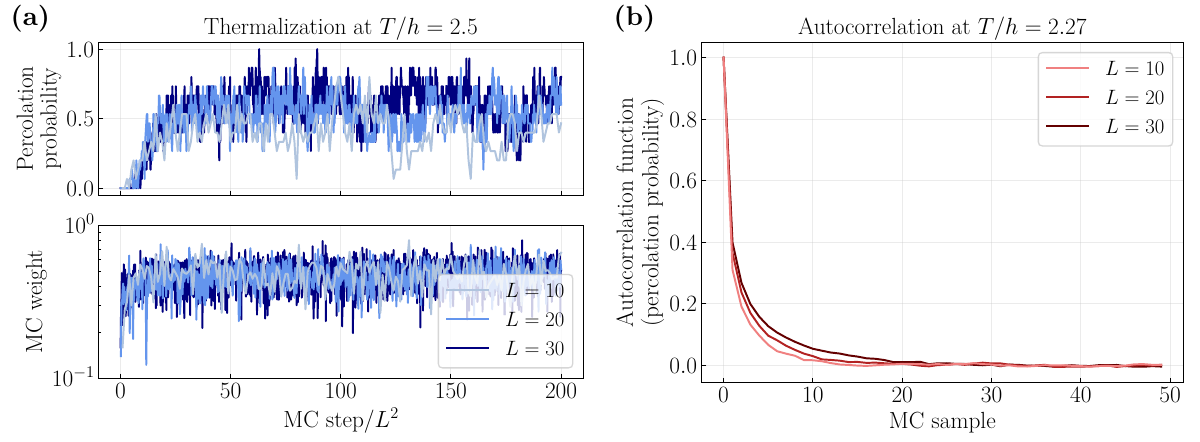}
\caption{\textbf{MC thermalization \& autocorrelation.} We show results for MC simulations of Hamiltonian~(\ref{eq:can_cl_ham}) for exactly two matter excitations on the square lattice. In panel~\textbf{(a)} we show the thermalization of the percolation probability (top) and the MC weights averaged over 15 runs (bottom) at $T/h=2.5$, i.e. slightly above the phase transition in the percolating phase. For clarity reasons, we only show every 100th measurement.  We have confirmed that in our simulations the system thermalizes after $200\!\times\!L^2$~steps for all~$T/h$. In panel~\textbf{(b)}, we show the autocorrelation of the percolation probability at $T/h=2.27$, i.e. approximately at the phase transition. We plot the average over 15 runs with~$10^4$ samples, respectively. Between each sample, we perform $2\!\times\!L^2$~steps. We find only small autocorrelation between samples.}
\label{fig:therm_auto}
\end{figure}

We thermalize the system with $100 \times L^2$ MC updates and perform $2 \times L^2$ MC updates between samples, see Fig.~\ref{fig:therm_auto}. We confirmed that the system is thermalized and the autocorrelation is low enough for every simulated parameter set. We take the autocorrelation between snapshots into account for every error bar.

For each sample, we measure the percolation probability, the percolation strength, the largest string cluster and the total number of strings in the system. For the case where we put only two matter particles into the system, we additionally measure the distance between those two matter particles. 

The algorithm to measure the percolation probability/strength in a system with periodic boundaries is presented in Alg.~\ref{alg:percolation}. The algorithm tries to find a path that winds around the system in at least one dimension, i.e. one can traverse the strings only visiting every string once and come back to the orginal lattice site while winding around the system at least once. In more technical terms, the algorithm starts a depth-first search from lattice sites in the starting point list $S$ while only traversing links that are strings. We assign a winding number to every vertex which changes when crossing over a cut through the system (in one dimension). When vertices with two distinct winding numbers (e.g. 0 and 1) ``meet'', the system is percolating in this specific dimension, since the path connecting the two vertices winds around the system and was thus assigned a different winding number. In the case of two or more dimensions, the percolation search is repeated for every dimension. The system is percolating when it is percolating in at least one dimension.

We use periodic systems only because the conventional percolation definition for open boundaries has an error that scales as $\mathcal{O}(1/L)$ in comparison to this definition where the error scales as $\mathcal{O}(1/L^2)$. This greatly improves the finite-size scaling.

\begin{figure}
\begin{algorithm}[H]
  \caption[Percolation detection (x-direction)]{\textbf{Percolation detection ($x$-direction).}}  \label{alg:percolation}
  \label{EPSA}
   \begin{algorithmic}[1]
\Require Graph representation of lattice with adjacency list $Adj$, starting point list $S$, $x$-coordinate list $x$, $x$-value $x_\mathrm{cut}$ which separates the system
\Ensure Percolation boolean (true $\equiv$ percolating; false $\equiv$ non-percolating)
\Function{IsPercolating}{$Adj, S, E$}
\State $\mathrm{discovered} \gets \{\mathrm{false}\}$ \Comment{No vertex is discovered in the beginning}
\For{$v_s \in S$}
        \State $\mathrm{winding\_number\_storage} \leftarrow \{\mathrm{INT\_MAX}\}$
        \State $\mathrm{v\_stack} \leftarrow \{\}$
        \State $\mathrm{winding\_number\_stack} \leftarrow \{\}$
        \State $\mathrm{top} \leftarrow 0$
        \State $\mathrm{v\_stack}[\mathrm{top}] \leftarrow v_s$ \Comment{Starting point of search}
        \State $\mathrm{winding\_number\_stack}[\mathrm{top}] \leftarrow 0$
        \While{$\mathrm{top} \geq 0$} 
        \State $v_\mathrm{top} \leftarrow \mathrm{v\_stack}[\mathrm{top}]$
        \State $w_\mathrm{top} \leftarrow \mathrm{winding\_number\_stack}[\mathrm{top}]$
        \State $\mathrm{top} \leftarrow \mathrm{top} - 1$
        \If{$\mathrm{discovered}[v_\mathrm{top}] = \mathrm{true}$}
        \State Continue with next iteration \Comment{Abandon search}
        \EndIf
        \State $\mathrm{discovered}[v_\mathrm{top}] \leftarrow \mathrm{true}$ 
        \State $\mathrm{winding\_number\_storage}[v_\mathrm{top}] \leftarrow w_\mathrm{top}$ 
        \For{$v_\mathrm{n} \in Adj[v_\mathrm{top}]$} 
        \If{string between $v_\mathrm{top}$ and $v_\mathrm{n}$}
        \State $w_\mathrm{new} \leftarrow w_\mathrm{top}$
        \If{$x[v_\mathrm{n}] > x_\mathrm{cut}$ \textbf{and} $x[v_\mathrm{top}] = x_\mathrm{cut}$}
        \State $w_\mathrm{new} \leftarrow w_\mathrm{top} + 1$
        \ElsIf{$x[v_\mathrm{top}] > x_\mathrm{cut}$ \textbf{and} $x[v_\mathrm{n}] = x_\mathrm{cut}$}
        \State $w_\mathrm{new} \leftarrow w_\mathrm{top} - 1$
        \EndIf
        \If{$\mathrm{discovered}[v_\mathrm{n}] = \mathrm{false}$}
        \State $\mathrm{top} \leftarrow \mathrm{top} + 1$
        \State $\mathrm{v\_stack}[\mathrm{top}] \leftarrow v_\mathrm{n}$
        \State $\mathrm{winding\_number\_stack}[\mathrm{top}] \leftarrow w_\mathrm{new}$
        \ElsIf{$\mathrm{winding\_number\_storage}[v_n] \neq w_\mathrm{new}$}
        \State \Return $\mathrm{true}$ \Comment{Algorithm has found percolating string}
        \EndIf
        \EndIf
        \EndFor
        \EndWhile
        \EndFor
        \State \Return $\mathrm{false}$ \Comment{Algorithm has not found percolating string}
\EndFunction
\end{algorithmic}
\end{algorithm}
\end{figure}

Since the percolation search is repeated many times (a typical number of samples is $\sim 10^5$) and we want to simulate as large systems as possible, the algorithm must be designed efficiently. In principle, a new depth-first search (see e.g. \cite{Even2011}) is performed for every starting point. Thus, the complexity would be $\mathcal{O}(D \times L^{D-1} \times L^D)$ since for every dimension we have $\mathcal{O}(L^{D-1})$ starting points from which we would start a depth-first search with complexity $\mathcal{O}(L^D)$. In practice, we can heavily reduce the complexity by storing visited lattice sites. In case we visit such a lattice site, the search stops. If we detect percolation, the search also stops.

For measuring the largest string cluster, we use a built-in function of the Boost C++ library that clusters the graph into its connected components only considering strings based on a depth-first search approach.

\subsection{Grand Canonical Monte Carlo Sampling} \label{sec:gc_sampling}

We simulate the classical Hamiltonian~(\ref{eq:gc_cl_ham})
\begin{align} 
    \hat{\mathcal{H}}_{\mathrm{gc}} = - h \sum_{l} \hat{\tau}^x_l - \mu \sum_{\j} \hat{n}_{\j} \nonumber
\end{align}
on the square and the cubic lattice. Since we are in a grand canonical ensemble, we fix the chemical potential $\mu$ in the system. The size of the lattice determines the matter density resolution that can be achieved by tuning the chemical potential.

Since we fix the gauge sector to $g_{\j} = +1$ and we only consider hard-core bosons, we can rewrite $\hat{n}_{\j}$ in terms of electric fields $\hat{\tau}^x_\ij$. For hard-core bosons, we have
\begin{align}
    \hat{n}_{\j} = \frac{1}{2} \, \bigl( 1 - (-1)^{\hat{n}_{\j}} \bigr).
\end{align}
Plugging in Gauss's law yields
\begin{align}
    \hat{n}_{\j} = \frac{1}{2} \, \biggl( 1 - \prod_{l \in +_{\j}} \hat{\tau}^x_l \biggr).
\end{align}
Finally, the grand canonical Hamiltonian reads
\begin{align} \label{eq:generalized_ising}
    \hat{\mathcal{H}}_{\mathrm{gc}} = - h \sum_{l} \hat{\tau}^x_l - \mu \, \sum_{\j} \, \frac{1}{2} \, \biggl( 1 - \prod_{l \in +_{\j}} \hat{\tau}^x_l \biggr).
\end{align}
To compare canonical results with matter particle number $N$ to grand canonical results, we have to tune $\mu$ and $\beta$ such that
\begin{align}
    N = \sum_{\j} \, \frac{1}{2} \, \biggl( 1 - \biggl\langle \prod_{l \in +_{\j}} \hat{\tau}^x_l \biggl\rangle \biggr).
\end{align}
Hamiltonian~(\ref{eq:generalized_ising}) does only depend on the electric field configuration $\{\hat{\tau}^x_l \}$ and is a generalized Ising model with four-spin-interactions $\propto \mu$ in a longitudinal magnetic field. 

The procedure for an MC update is much simpler than in the canonical case. In the first update procedure (which can change the number of matter particles), a random link is chosen and we propose a flip to that link. If the move is accepted, the electric field on the link is flipped and the matter particle numbers at the neighboring lattice sites are updated to satisfy Gauss's law. As the canonical plaquette update does not change the number of matter particles, it can be used here as well. Neither of the two updates has asymmetric TPDs and thus we can use Metropolis sampling. In every MC update, there is a 50\% chance to propose either a plaquette update or an electric field flip update.

We thermalize the system with $500 \times L^2$ MC updates and perform $5 \times L^2$ MC updates between samples. We confirmed that the system is thermalized and the autocorrelation is low enough for every simulated parameter set. We take the autocorrelation between snapshots into account for every error bar.

As in the canonical case, we measure the percolation probability, the percolation strength, the largest string cluster and the total number of strings in the system for each sample. We use the same algorithms as in the canonical case.

\section{(Quantum) Monte Carlo results}
\label{app:quantum_tc}

Here we provide supplementary results supporting the claims in the main part of the paper, and details on the procedure used to extract the critical exponents. 

\subsection{Classical finite-density regime on the square lattice}
\label{app:classical_finite_matter_density}

\begin{figure}[t]
\includegraphics[width=\textwidth]{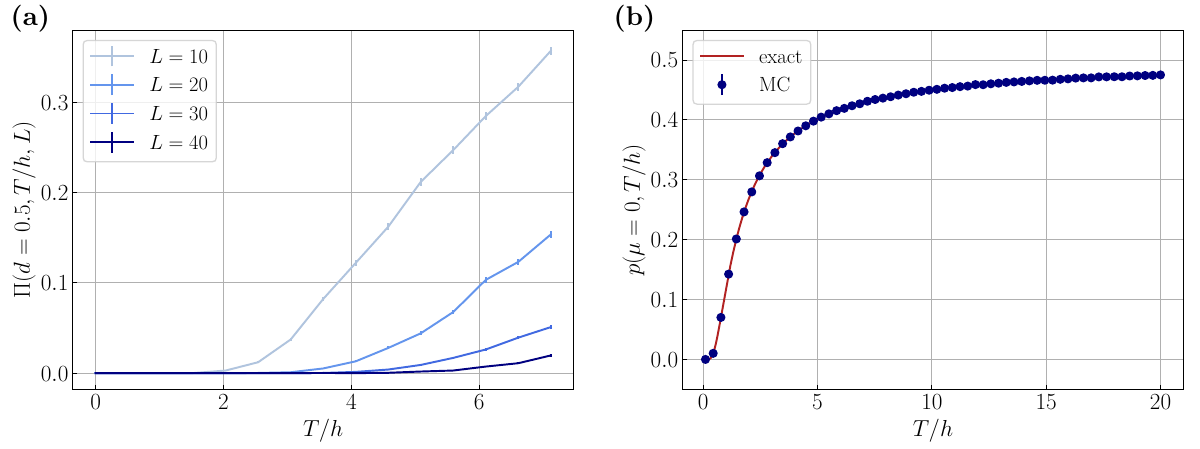}
\caption{\textbf{Finite-density regime.} We demonstrate the non-existence of thermal deconfinement at non-zero matter density. In panel~\textbf{(a)} we show results for Hamiltonian~(\ref{eq:can_cl_ham}) at density $d=0.5$. For increasing system size, the percolation probability $\Pi$ decreases in magnitude and the first non-zero shifts to higher temperatures $T/h$. In panel~\textbf{(b)}, we show results for Hamiltonian~(\ref{eq:gc_cl_ham}) at $\mu=0$. The error bars are too small to be visible. With increasing $T/h$, the probability $p$ for a bond to host a string approaches the percolation threshold $p_\mathrm{c,square} = 0.5$ from below confirming the exact result $p = e^{-\beta h} / 2 \cosh(\beta h)$. Because the percolation threshold is only reached at $T/h \to \infty$, non-zero percolation at $\mu=0$ in finite systems is only a finite-size effect and there is no thermal deconfinement phase transition.}
\label{fig:finite_matter_density}
\end{figure}

We simulate Hamiltonian~(\ref{eq:can_cl_ham}) at non-zero matter density $d=0.5$ for $L^2 = 10^2,20^2,30^2,40^2$ and show the percolation probability $\Pi(T)$ in Fig.~\ref{fig:finite_matter_density}a. While $\Pi$ is non-zero at high temperatures for every simulated finite-size system, we observe a decrease in $\Pi$ with increasing system size.

At $\mu=0$, the electric fields are completely independent and the probability $p$ for a bond to host a string is given by $p = e^{-\beta h} / 2 \cosh(\beta h)$, i.e. the percolation threshold $p_\mathrm{c,square} = 0.5$ is never reached for finite temperatures. We simulate Hamiltonian~(\ref{eq:gc_cl_ham}) at $\mu=0$ and show $p(T/h)$ as a function of temperature $T/h$ in Fig.~\ref{fig:finite_matter_density}b. $p$ approaches $0.5$ from below for $T/h \to \infty$. Hence, we proved that a thermal deconfinement phase transition cannot exist for finite $T/h$ at $\mu=0$. We conjecture the absence of a phase transition for every non-zero matter density, as indicated by our numerical simulations; we performed T-sweeps for $d=0.1,0.2,...,1.0$ and find the same qualitative behavior as in Fig.~\ref{fig:finite_matter_density}a for all simulated densities.

\begin{figure}[t]
\includegraphics[width=\textwidth]{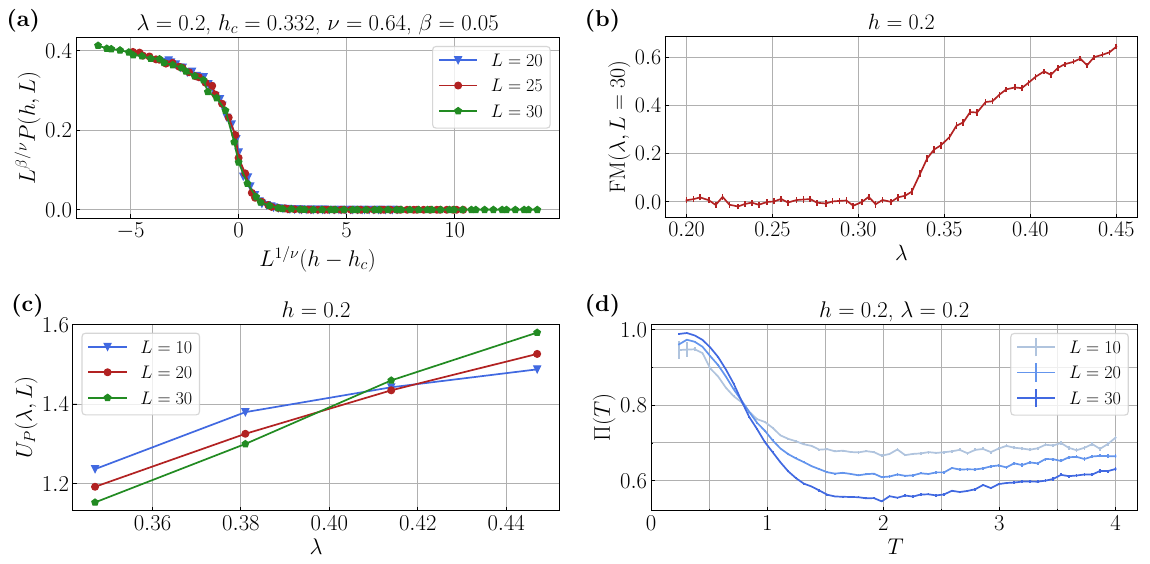}
\caption{\textbf{QMC finite-size scaling.} We show results for the POPs in the square lattice toric code Hamiltonian~(\ref{eq:eTC}). In panel~\textbf{(a)} we present the finite-size data collapse of the percolation strength $P$ for $\lambda=0.2$. The curves collapse into a well-defined scaling function. In panel~\textbf{(b)} we present the Fredenhagen-Marcu order parameter $\mathrm{FM}(\lambda, L=30)$ -- obtained from QMC simulations in the $\hat{\tau}^z$-basis -- for $h=0.2$, confirming the known phase boundary at $\lambda_c \sim 0.33$. In panel~\textbf{(c)}, we present the percolation strength Binder cumulant $U_P(\lambda)$ for different system sizes. We observe a shift in the crossing points toward lower $\lambda$ which is a clear sign of strong finite-size effects. We conjecture that the crossing points approach the known critical value $\lambda_c \sim 0.33$ in the TDL. In panel~\textbf{(d)}, we present the percolation probability $\Pi(T)$ for different system sizes. At temperatures well above the bulk gap, we observe a clear decrease in $\Pi$ with higher system sizes. For Bernoulli percolation, Kolmogorov’s zero-one-law \cite{Lyons1990} restricts the percolation probability to either zero or one in the TDL. We conjecture that $\Pi$ approaches zero in the TDL for any $T>0$.}
\label{fig:qmc_demonstration}
\end{figure}

\subsection{Extended toric code QMC}
\label{app:qmc}

We simulate Hamiltonian~(\ref{eq:eTC}) on a periodic square lattice using an adaption of the continuous-time QMC algorithm from Wu~\textit{et al.}~\cite{Wu2012}. Unless explicitly stated otherwise, we sample snapshots in the $\hat{\tau}^x$-basis. We set $\mu=J=1$.

\textit{$h$-sweeps.---} We set $T = 1/L$, $\lambda=0,0.1,0.2,0.3$ and $L^2=20^2,25^2,30^2$. To extract the critical electric field $h_c$ and the critical exponents $\nu$ (correlation length) and $\beta$ (percolation strength $P$) we first perform a crossing point analysis of the percolation strength Binder cumulant
\begin{align}
    U_P(h, L) = \frac{\langle P^4(h, L) \rangle}{\langle P^2(h, L) \rangle^2}
\end{align}
to extract an estimate for $h_c$. We obtain an estimate for $\nu$ by manually collapsing finite-size curves using the scaling ansatz
\begin{align}
    U_P(h, L) = f_{U_P} \bigl( L^{1/\nu} (h-h_c) \bigr),
\end{align}
where $f$ is the scaling function. Finally, $\beta$ can be estimated by manually collapsing finite-size curves with the scaling ansatz
\begin{align}
    P(h, L) = L^{-\beta/\nu} f_{P} \bigl( L^{1/\nu} (h-h_c) \bigr),
\end{align}
see Fig.~\ref{fig:qmc_demonstration}a. Our estimates serve as the starting point for a reduced $\chi^2$ optimization which converged to the final values $h_c=0.3301(17)$, $\nu=0.65(27)$ and $\beta=0.05(4)$ for $\lambda = 0$. The critical exponents are identical for different $\lambda$ up to statistical errors as expected. We were unable to provide a good estimate of the Fredenhagen-Marcu order parameter in the confined phase for $h > \lambda$ due to high noise in the data.

\textit{$\lambda$-sweep.---} We set $T = 1/L$, $h=0.2$ and $L^2=10^2,20^2,30^2$. We find Binder cumulant crossing points around $\lambda \approx 0.4$ that shift toward lower $\lambda$ with increasing system size, see Fig.~\ref{fig:qmc_demonstration}c. The gap consists of the bulk gap $\Delta_\mathrm{B}$ and the finite-size gap $\Delta_L$. Because of the self-duality, we know that the true phase transition where the gap closes is at $\lambda_c \sim 0.33$. However, in finite-size systems (i) the gap never completely closes and (ii) the finite-size ground state can \textit{appear} topological because of the finite-size gap. We conjecture that the true critical value of $\lambda_c \sim 0.33$ is reached in the TDL and the strong finite-size effects are caused by a large finite-size gap. In Fig.~\ref{fig:qmc_demonstration}b, we show the Fredenhagen-Marcu order parameter
\begin{align}
    \mathrm{FM} = \frac{\prod_{l \in \mathcal{C}_{1/2}} \hat{\tau}^z_l}{\prod_{l \in \mathcal{C}} \hat{\tau}^z_l},
\end{align}
where $\mathcal{C}$ is a closed contour of links with length $\sim L$ at equal imaginary time and $\mathcal{C}_{1/2}$ contains half the links of $\mathcal{C}$ and thus forms an open contour. $\mathrm{FM}$ was calculated using QMC snapshots in the $\hat{\tau}^z$-basis and confirms the known critical value $\lambda_c \sim 0.33$.

\begin{figure}[t]
\includegraphics[width=\textwidth]{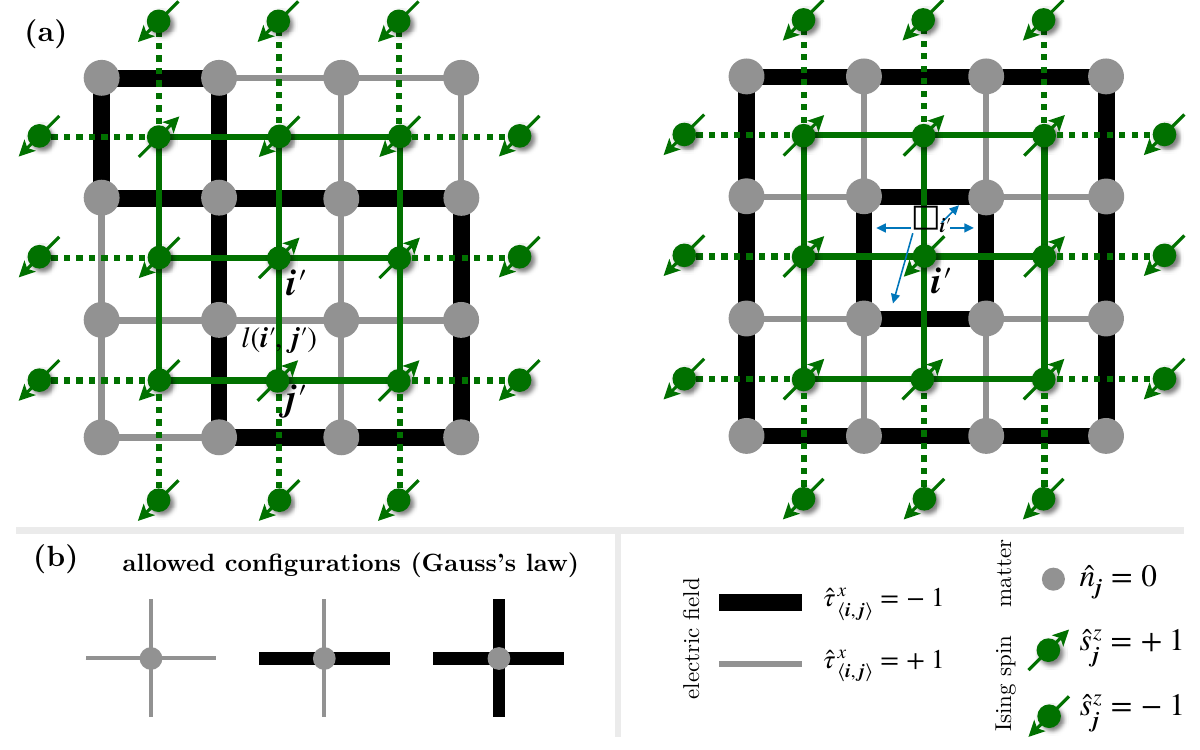}
\caption{\textbf{Ising mapping of the pure gauge model.} We illustrate the mapping of the pure gauge model Hamiltonian~(\ref{eq:pure_gauge_model}) to the TFIM on the dual lattice. In panel~\textbf{(a)} we show two configurations of electric strings and the corresponding spin configuration on the dual lattice. The electric strings form closed loops. The electric field on a link $l$ depends on the dual lattice link between $\i^\prime$ and $\j^\prime$ that intersects $l$. By flipping a spin on a dual lattice site $\j^\prime$, all electric fields in the surrounding plaquette $\square_{\j^\prime}$ are flipped. A percolating string corresponds to a domain wall of spins since closed string loops necessarily have spins of the same orientation along their perimeter. Due to Gauss's law, the star term $\propto \mu$ is constant, see panel~\textbf{(b)}. Adapted from \cite{Wegner2014}.}
\label{fig:ising_mapping}
\end{figure}

\textit{$T$-sweep.---} We set $h=\lambda=0.2$ and $L^2=10^2,20^2,30^2$. We find a crossing-point in the percolation strength Binder cumulant at $T>0$ and a percolating phase extending into $T>0$, see Fig.~\ref{fig:qmc_demonstration}d. However, the topological order is known to break down for $T>0$. We trace this behavior back to the bulk gap $\Delta_\mathrm{B}$. For $T<\Delta_\mathrm{B}$ the density of thermal excitations is non-zero in the TDL, however, it is exponentially suppressed and thus not visible in finite-size systems. After all, this is the reason why we can gain insights into the ground state using finite-$T$ QMC schemes. While the true ground state for $T>0$ is not topological, it \textit{appears} like the topological ground state at $T=0$ in finite-size systems. This reasoning is supported by the estimated order-of-magnitude of the bulk gap $\Delta_\mathrm{B} \simeq J$ at $h=\lambda=0.2$ which matches the temperature scale below which we observe the percolating (deconfined) ground state features. We were unable to provide a good estimate of the Fredenhagen-Marcu order parameter for finite temperatures due to high noise in the data.

\section{Mapping the pure gauge theory to the 2D transverse-field Ising model}
\label{app:ising_mapping}

Here we concisely motivate the mapping of the pure gauge extended toric code Hamiltonian~(\ref{eq:eTC}), where $\lambda=0$ (i.e. without matter), to the 2D transverse-field Ising model (TFIM) \cite{Wegner1971}. The starting point is the pure gauge model
\begin{align} \label{eq:pure_gauge_model}
    \hat{\mathcal{H}} = &- \mu \sum_+ \prod_{l \in +} \hat{\tau}_l^x \; - J \sum_{\square} \prod_{l \in \square} \hat{\tau}_l^z - h \sum_{l} \hat{\tau}_l^x \;.
\end{align}
Since we have no matter, the \Ztwo{} electric strings can only form closed loops. If we define Ising spins $\hat{s}^z \in \{\pm 1\}$ on the dual lattice sites, a string loop on an elementary plaquette can be thought of as one spin with $\hat{s}^z=+1$ in a background of spins with $\hat{s}^z=-1$. The operator $\hat{s}^x$ flips the spin $\hat{s}^z$ in the dual lattice and thus the electric fields on the surrounding plaquette in the LGT: it creates/destroys string loops. This intuition can be formalized with the definition of the dual-lattice site variables:
\begin{align}
    \hat{s}^z_{\i^\prime} \hat{s}^z_{\j^\prime} &= \hat{\tau}^x_{l(\i^\prime, \j^\prime)}, \\
    \hat{s}^x_{\j^\prime} &= \prod_{l \in \square_{\j^\prime}} \hat{\tau}_l^z,
\end{align}
where $l(\i^\prime, \j^\prime)$ is the link that intersects with the dual link between dual lattice sites $\i^\prime, \j^\prime$ and $\square_{\j^\prime}$ is the plaquette of links surrounding the dual lattice site $\j^\prime$, see Fig.~\ref{fig:ising_mapping}a.
Plugging into Hamiltonian~(\ref{eq:pure_gauge_model}) yields the TFIM:
\begin{align}
    \hat{\mathcal{H}}^\prime_{\mathrm{TFIM}} = - h \sum_{\ijprime} \hat{s}^z_{\i^\prime} \hat{s}^z_{\j^\prime} - J \sum_{\i^\prime} \hat{s}^x_{\i^\prime}.
\end{align}
The star term $\propto \mu$ vanishes due to Gauss's law, see Fig.~\ref{fig:ising_mapping}b. The square lattice TFIM has a quantum critical point at $(h/J)_c \approx 1/3.044 \approx 0.33$ \cite{Croo1998, Blote2002} (3D Ising universality) and a critical temperature $(T/h)_c = 2 / \log(1+\sqrt{2}) \approx 2.27$ \cite{Onsager1944} (2D Ising universality). We identify the (non-)percolating phase of the extended toric code with the unmagnetized (magnetized) phase of the TFIM. Percolating strings in the deconfined phase correspond to domain walls in the paramagnetic phase in the dual TFIM since closed string loops necessarily have spins of the same orientation along their perimeter.

\end{document}